\documentclass[preprints,perspective,accept,pdftex,moreauthors]{Definitions/mdpi} 

\firstpage{1} 
\pubvolume{1}
\issuenum{1}
\articlenumber{0}
\pubyear{2025}
\copyrightyear{2025}

\datereceived{ } 
\daterevised{ } 
\dateaccepted{ } 
\datepublished{ } 

\hreflink{https://doi.org/}

\Title{Axion Searches at the CERN SPS: From Their Dawn to Current Prospects}

\TitleCitation{Axion Searches at the CERN SPS: From Their Dawn to Current Prospects}

\Author{Paolo Crivelli$^{1}$\orcidA{}*, Martina Mongillo\orcidB{}$^{1}$ }

\AuthorNames{Paolo Crivelli}

\isAPAStyle{%
       \AuthorCitation{Lastname, F., Lastname, F., \& Lastname, F.}
         }{%
        \isChicagoStyle{%
        \AuthorCitation{Lastname, Firstname, Firstname Lastname, and Firstname Lastname.}
        }{
        \AuthorCitation{Lastname, F.; Lastname, F.; Lastname, F.}
        }
}

\address{%
$^{1}$ \quad ETH Zurich, Institute for Astrophysics and Particle Physics, 8093 Zurich, Switzerland.}

\corres{Correspondence: crivelli@phys.ethz.ch}

\abstract{
This mini-review traces the evolution of axion searches at the CERN Super Proton Synchrotron (SPS), beginning with the early proposal by Guido Barbiellini in 1982 and culminating in the recent advances of the NA62 and NA64 experiments. We discuss the experimental strategies employed in early beam dump searches, the current status of axion and axion-like particle (ALPs) searches at the CERN SPS and future directions. This review serves as a tribute to Guido Barbiellini’s scientific legacy and his visionary contributions to this field.
}

\begin{document}

\section{Introduction}
The search for axions and axion-like particles (ALPs) is a very active field trying to address longstanding puzzles in particle physics~\cite{Sikivie:2006ni,Jaeckel:2010ni,IRASTORZA201889}. Originally proposed as a consequence of the Peccei–Quinn mechanism~\cite{Peccei:1977hh} to solve the strong CP problem~\cite{Wilczek:1977pj, Weinberg:1977ma}, axions are also viable candidates for light, weakly interacting bosons that could account for dark matter (see e.g. \cite{Adams:2022pbo} for a recent review).

While collider experiments have largely explored the high-energy frontier, fixed-target setups offer distinct advantages in probing light, weakly coupled Dark Sectors in the MeV-GeV mass range \cite{PBC:2025sny}. These configurations are also well suited for investigating long-lived axions and ALPs that might evade detection in traditional collider environments.

Among various experimental strategies, fixed-target facilities such as the North Area at the CERN Super Proton Synchrotron (SPS) offer a powerful and flexible tool to investigate such elusive particles, owing to their high-intensity and high-energy beams and the possibility to employ adaptable experimental configurations~\cite{Banerjee:2774716}.

A significant experimental gap exists between collider-based searches and traditional beam dump approaches \cite{cernnews}. In this intermediate domain, where new particles may be too weakly coupled for collider detection yet too short-lived for conventional beam dump strategies, experiments like NA64 play a crucial role. By combining high-intensity electron and muon beams with carefully designed and instrumented detectors, NA64 is capable of probing rare processes with exceptional sensitivity~\cite{NA64:2025ddk}.

This mini-review aims to trace the evolution of axion searches at the CERN SPS, from Guido Barbiellini’s early proposals to the modern results obtained by NA62 and NA64. 

\section{Early Beam Dump Searches for Axion-Like Particles}

The experimental pursuit of axions and ALPs dates back to the early 1980s, when several beam dump experiments laid the groundwork for future searches. These efforts were instrumental in developing the conceptual and methodological tools that underpin modern investigations.

\subsection{SLAC E137: An Electron Beam Dump Pioneer}
Between 1980 and 1982, the E137 experiment at SLAC employed a 20 GeV electron beam incident on an aluminium target. The goal was to search for long-lived neutral particles that could penetrate a thick shielding region and decay into detectable products downstream. The configuration, which included a 179-meter shield followed by a decay volume and detector, allowed sensitivity to particles like ALPs decaying predominantly into two photons or in electron-positron pairs \cite{Bjorken:1988as}.

\subsection{Guido Barbiellini’s Proposal at the SPS}
At CERN, Guido Barbiellini advanced the idea of searching for axions in a proton beam dump configuration. In his 1982 contribution to the Workshop on SPS Fixed Target Physics for the Years 1984–1989 \cite{barbiellini1982sps}, he emphasised the potential of detecting axions produced by high-energy proton collisions with a fixed target through their decay into photon/lepton pairs within the CHARM detector volume \cite{CHARM:1985anb}. This proposal was among the first to explicitly call for axion searches in such a context.
Guido outlined a testable strategy at a time when axion searches were far from mainstream and much of the focus in experimental high-energy physics to search for "new phenomena and small deviation from orthodoxy"\footnote{Searches for BSM physics in the words of Guido \cite{barbiellini1982sps}.} was shifting toward supersymmetry and heavy neutrinos.

This strategy naturally pointed to the CHARM experiment as a suitable setup: a high-intensity beam dump configuration originally designed for the study of prompt neutrino interactions and rare decays. In CHARM, 400 GeV protons from the SPS were directed onto a thick target, generating a flux of secondary particles, including mesons, that decayed within a shielded decay tunnel. Downstream,  calorimeters and tracking chambers were used to detect the decay products of long-lived neutral particles.

What made CHARM an effective probe for axion searches was its long decay region, large fiducial volume, and shielding that suppressed Standard Model backgrounds. 
The underlying idea was that axions ($a$) could be produced through proton-nucleus interactions~\cite{Donnelly:1978ty}\footnote{More recently, the CHARM results were reinterpreted including the possibility for meson decays (e.g., $\pi^0, \eta \to \gamma a$)~\cite{Dobrich:2019dxc}.}, and subsequently decay into two photons ($a \to \gamma \gamma$) or in lepton pairs ($a \to e^+e^-,a \to \mu^+\mu^-$) within the decay volume~\cite{CHARM:1985anb}. 
The signature would be penetrating neutral particles, without accompanying tracks, and with energy deposition consistent with an electro-magnetic shower.

This proposal directly influenced the CHARM collaboration’s later analysis, which set one of the first experimental constraints on axion-like particles with photon couplings in the MeV to GeV mass range. In their 1986 publication \cite{CHARM:1985anb}, the CHARM collaboration presented limits on long-lived, weakly interacting neutral particles decaying into two photons or a pair of leptons, using the type of beam dump and decay volume configuration Guido had envisioned.

\subsection{Other Early Contributions}
Additional notable pioneering experiments include the SLAC E141~\cite{Riordan:1987aw} and Fermilab E774~\cite{Bross:1989mp}, both of which probed ALP decay modes involving photon and electron pairs; the NA3 experiment at CERN~\cite{NA3:1986ahv}, which used pion beams to explore similar phenomena; and NuCal~\cite{Blumlein:1990ay}, which searched for ALPs in a proton beam dump configuration; and NOMAD, which set constraints on ALPs with eV-scale masses using the SPS wide-band neutrino beam at CERN~\cite{Gninenko:2000ds}.

\section{Modern Axion Searches at the CERN SPS: NA62 and NA64}
\begin{figure}[h!]
\centering
\includegraphics[width=\columnwidth]{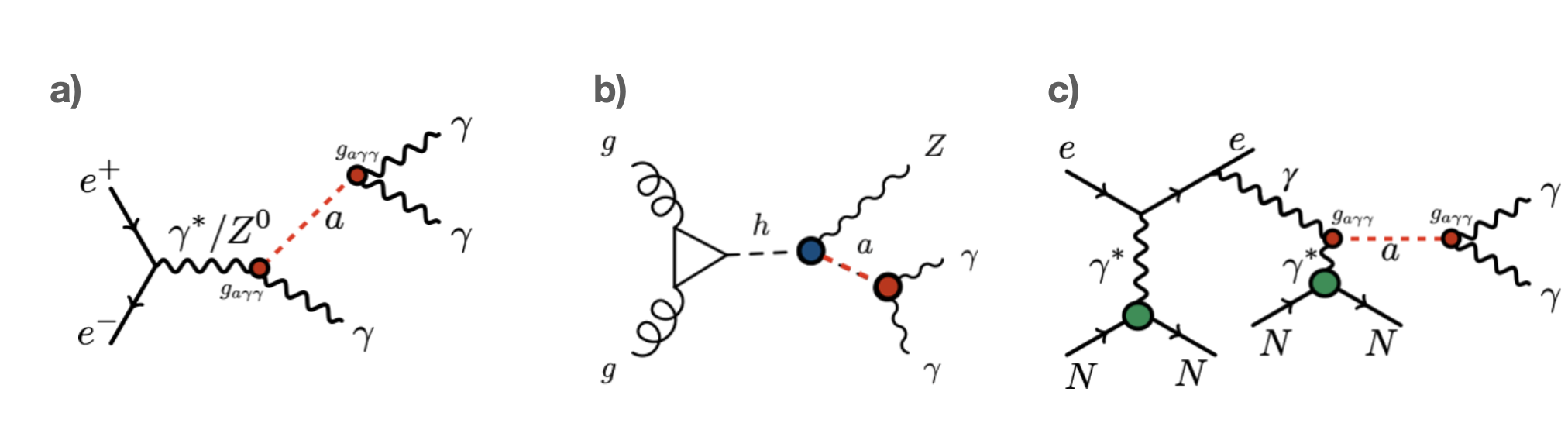}
\caption{\small{a) Production of ALPs at e+e- colliders. At LEP, searches were focused at the $Z_0$ pole, while at BELLE II, ALPs are produced via virtual annihilation photon coupling to the ALPs. The signature is either a monophoton (if the ALP decays outside the detector) or a 3-photon final state (if it decays within the detector). b) Example of ALP production at LHC. c) Production of ALPs in an electron beam dump experiment. High-energy (100 GeV) electrons impinging on the dump emit Bremsstrahlung photons, which can produce ALPs via the Primakoff effect in the nuclear field.}} 
\label{fig:ALPsProduction}
\end{figure}

NA64 represents a new generation fixed-target experiment optimised for the search for Dark Sectors~\cite{Gninenko:2013rka, Gninenko:2016kpg}. It utilises $\sim$100 GeV electron~\cite{NA64:2023wbi}, positron~\cite{NA64:2023ehh,NA64:2025rib}, hadron~\cite{NA64:2024mah} and muon~\cite{NA64:2024klw,NA64:2024nwj} beams from the CERN SPS, which are directed onto an active target consisting of an electromagnetic calorimeter (ECAL) (for an overview of the experiment, see Ref.~\cite{NA64:2025ddk}). Axions or ALPs can be produced via the interaction of electrons with the electromagnetic field of the target nuclei via the Primakoff process (see Fig.~\ref{fig:ALPsProduction}c)~\cite{Dusaev:2020gxi}.

\begin{figure}[h!]
\centering
\includegraphics[width=1.\columnwidth, trim=200 20 750 250, clip]{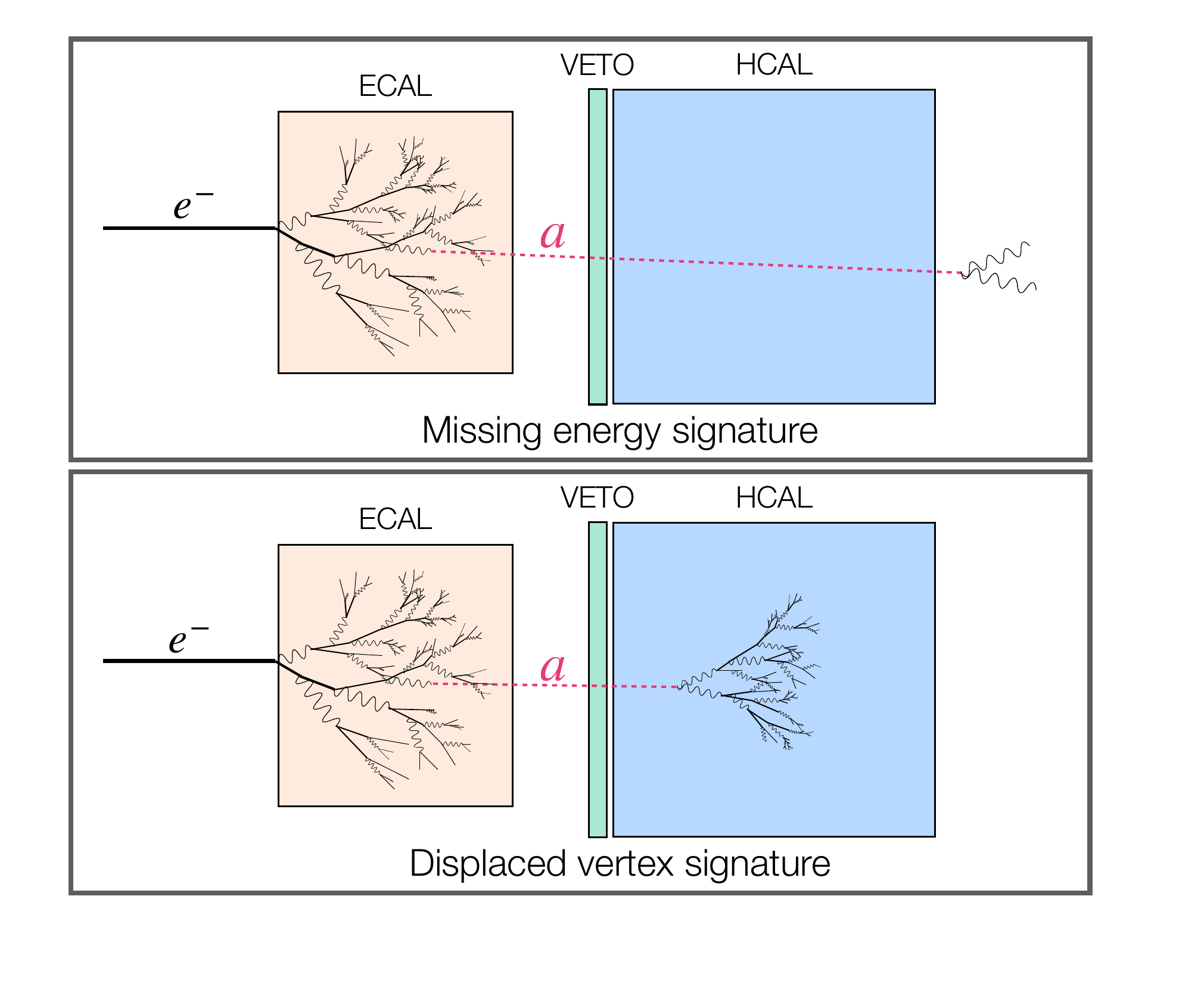}
\caption{\small{Schematic illustration of two possible ALP signatures in the NA64 setup. The ALP is produced via the reaction chain $e^-Z \to e^-Z\gamma$ followed by $\gamma Z \to aZ$, initiated by 100 GeV electrons interacting in the active electromagnetic calorimeter (ECAL), where $Z$ denotes the nuclei of the target material.
Top: in the case of a long-lived ALP, the particle escapes the detector before decaying, resulting in a missing energy signal.
Bottom: for a shorter-lived ALP, the decay $a \to \gamma\gamma$ can occur inside the hadronic calorimeter (HCAL), producing a displaced vertex signal.}}
\label{fig:NA64_sketch}
\end{figure}

If produced, ALPs may either escape the detector before decaying, resulting in a distinctive missing-energy signature, or decay within the detector volume, appearing as a displaced vertex. Both signatures are illustrated in Fig.\ref{fig:NA64_sketch}. 

Besides the core of the experiment, the target ECAL, the NA64 detector comprises several subsystems designed to suppress backgrounds and enhance signal sensitivity. A synchrotron radiation detector is used to tag incoming electrons, while veto counters and hadronic calorimeters (HCALs) screen for SM backgrounds. For displaced vertex signatures, the VETO combined with the first HCAL module play a crucial role in suppressing backgrounds from hadronic secondaries produced in the target. Additionally, the HCAL's transverse segmentation is essential for distinguishing purely electromagnetic signal-like showers. High-precision tracking chambers ensure accurate reconstruction of the incoming particle momentum~\cite{NA64:2023wbi}.

\begin{figure}[h!]
\centering
\includegraphics[width=1\columnwidth]{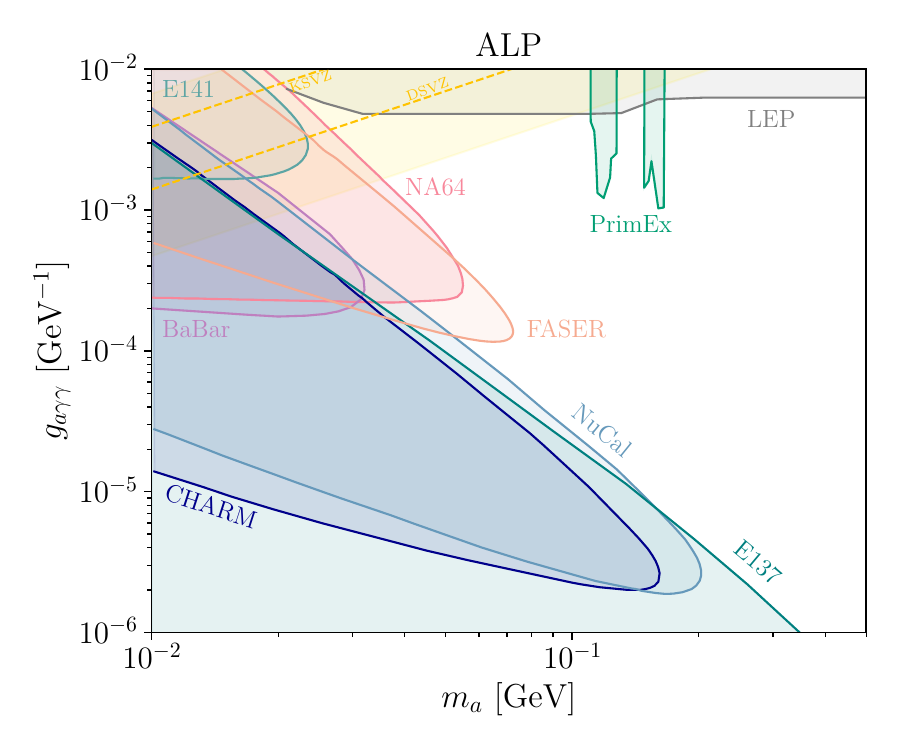}
\caption{\small{Current bounds for ALPs coupling predominantly to photons in the ($m_a; g_{a\gamma\gamma}$)- plane as a function of the (pseudo)scalar mass $m_a$.  The yellow band represents the parameter space for the benchmark QCD axion (DFSZ~\cite{Dine:1981rt} and KSVZ~\cite{Kim:1979if} models extended with a broader range of E/N values~\cite{ParticleDataGroup:2018ovx,Kim:1998va}).}}
\label{fig:exclusionPlot}
\end{figure}

In 2020, the NA64 collaboration published its first results on ALP searches, setting world-leading limits on the ALP-photon coupling in the mass range below 100 MeV. Both signatures—missing energy from ALPs escaping the detector and displaced vertices from ALPs decaying within it—were investigated in the analysis. Based on a dataset of  \(2.84 \times 10^{11}\) electrons on target, no excess events consistent with either scenario were observed. The resulting exclusion limits began to close the gap between collider-based constraints and those from previous beam dump experiments \cite{NA64:2020qwq}, as illustrated by the rose-shaded region in Fig.~\ref{fig:exclusionPlot}.

In parallel, the NA62 experiment has also contributed to the search for axion-like particles, particularly via rare kaon decay channels meson decays \cite{Dobrich:2019dxc}. NA62 is also a fixed-target experiment located in CERN's North Area, primarily designed to measure the ultra-rare decay $K^+ \to \pi^+ \nu \bar{\nu}$ using a 400 GeV proton beam extracted from the SPS~\cite{Anelli:2005ju}. Secondary 75 GeV positively charged hadrons, mostly kaons, are transported along a 100-meter beamline to the detector. The experimental apparatus includes a differential Cherenkov counter for kaon identification, a straw tracker in vacuum, a ring-imaging Cherenkov detector, and a system of calorimeters for photon and muon vetoing~\cite{NA62:2017rwk}. These features provide excellent momentum resolution and timing, making NA62 particularly well suited to search for rare kaon decays involving light new particles such as ALPs.
NA62 searched for ALP coupling predominantly to gluons and to fermions (see e.g.~\cite{Beacham:2019nyx}), produced via the decay \( K^+ \rightarrow \pi^+ a \), where the ALP can then decay respectively to leptons or hadrons~\cite{NA62:2025yzs}. Limits were placed also on invisible ALP looking for \(\pi^{0}\rightarrow inv\) events~\cite{NA62:2020pwi} or via \( K^+ \rightarrow \pi^+ inv \)~\cite{NA62:2020xlg}.
Using 2017–2018 data, NA62 also searched for gluon-coupled ALPs looking for the decay chain \( K^+ \rightarrow \pi^+ a, a\rightarrow\gamma \gamma \), where the ALP decays promptly to two photons~\cite{NA62:2023olg}. No signal was observed, and upper limits were set on the product of branching ratios \( \text{BR}(K^+ \rightarrow \pi^+ a) \times \text{BR}(a \rightarrow \gamma\gamma) \), as a function of the ALP mass. 

\section{Future Prospects at CERN}
NA64 has already recorded an order of magnitude more data compared to the published results, and plans to collect up to $10^{13}$ electrons on target after LS3. This represents an increase by a factor $\sim$40 in statistics compared to the dataset used in the 2020 ALP analysis, which will boost the experiment’s sensitivity towards the unexplored area at higher masses and lower couplings.

However, in the high-coupling region, ALPs decay inside the ECAL and cannot be distinguished from a SM electromagnetic shower. Since this loss is exponential, increasing statistic only leads to a marginal improvement in sensitivity. Nevertheless, this region remains theoretically compelling, especially within the QCD axion band (yellow region in Fig.~\ref{fig:exclusionPlot}).

Three strategies are under consideration to probe this interesting parameter space. The first one involves shortening the dump by employing a more compact target calorimeter placed in front of the ECAL, with the ECAL then used to identify visible decays into photon pairs. The second strategy proposes a modifications to the invisible mode setup by shortening the first HCAL module used as VETO for electro-nuclear events, or further segmenting the HCAL to improve discrimination between hadronic and electromagnetic showers from diphoton ALP decays. These approaches build on insights from previous studies, including NA64's visible search for the X17 boson~\cite{NA64:2019auh, NA64:2020xxh}. A third possibility involves upgrades on the NA64 muon-mode setup, including the modification to the HCAL central cell segmentation, to enable detection of two separate electromagnetic showers from ALP decays. 

Complementing the ALP search with a muon active dump, as in NA64$\mu$, could also enable the exploration of previously inaccessible regions of the ($m_a; g_{a\gamma\gamma}$) parameter space~\cite{Li:2025yzb}.
In addition to the photon-coupled model, leptophilic ALPs represent another interesting scenario that NA64 plans to explore in future studies~\cite{Eberhart:2025lyu,Li:2025yzb}.

NA62 is currently operating in a beam-dump mode, with new results expected on ALPs. The projections from the full expected 2021-2026 dataset are anticipated in Ref.~\cite{Jerhot:2936260}. 

The FASER experiment at LHC has recently released their first results in the search for ALPs~\cite{FASER:2024bbl}. Located 480 meters downstream of the ATLAS interaction point~\cite{Feng:2017uoz}, FASER explores a similar region of parameter space, targeting light, long-lived particles. Its sensitivity complements that of NA64, particularly in the regime of smaller couplings and higher masses.

Looking further ahead, the SHiP experiment, currently under construction at the CERN SPS, aims to probe heavier masses and smaller couplings~\cite{SHiP:2021nfo}. SHiP plans to use a dedicated beam dump and a long decay volume to search for a broad range of hidden sector particles, including ALPs. It is complementary to NA62, NA64 and FASER, filling sensitivity gaps where these experiments' reach decreases.

\section{Conclusion}
The search for axions and axion-like particles has evolved into one of the most compelling frontiers in the quest to uncover physics beyond the Standard Model. At CERN, the SPS is continuing to serve as a versatile platform for such searches. More than four decades ago, Guido Barbiellini proposed using the SPS beam dump configuration to investigate the existence of axions. Today, new experiments continue this inquiry, building on Guido’s original insight. Together, NA64, NA62, FASER and SHiP efforts represent a comprehensive, multi-prong approach to explore the ALP parameter space in the MeV-GeV region at CERN.
We conclude quoting Guido's still very valid recommendation to physicists willing to search for small effects not predicted by the Standard Model \cite{barbiellini1982sps}: "Try to be a little bit heretical (but not too much!)".

\bibliographystyle{unsrt}
\bibliography{references}

\begin{thebibliography}{999}

\bibitem[Sikivie(2008)]{Sikivie:2006ni}
Sikivie, P.
\newblock {Axion Cosmology}.
\newblock {\em Lect. Notes Phys.} {\bf 2008}, {\em 741},~19--50,
  \href{http://arxiv.org/abs/astro-ph/0610440}{{\normalfont
  [astro-ph/0610440]}}.
\newblock {\url{https://doi.org/10.1007/978-3-540-73518-2_2}}.

\bibitem[Jaeckel and Ringwald(2010)]{Jaeckel:2010ni}
Jaeckel, J.; Ringwald, A.
\newblock {The Low-Energy Frontier of Particle Physics}.
\newblock {\em Ann. Rev. Nucl. Part. Sci.} {\bf 2010}, {\em 60},~405--437,
  \href{http://arxiv.org/abs/1002.0329}{{\normalfont
  [arXiv:hep-ph/1002.0329]}}.
\newblock {\url{https://doi.org/10.1146/annurev.nucl.012809.104433}}.

\bibitem[Irastorza and Redondo(2018)]{IRASTORZA201889}
Irastorza, I.G.; Redondo, J.
\newblock New experimental approaches in the search for axion-like particles.
\newblock {\em Progress in Particle and Nuclear Physics} {\bf 2018}, {\em
  102},~89--159.
\newblock {\url{https://doi.org/https://doi.org/10.1016/j.ppnp.2018.05.003}}.

\bibitem[Peccei and Quinn(1977)]{Peccei:1977hh}
Peccei, R.D.; Quinn, H.R.
\newblock {CP Conservation in the Presence of Instantons}.
\newblock {\em Phys. Rev. Lett.} {\bf 1977}, {\em 38},~1440--1443.
\newblock {\url{https://doi.org/10.1103/PhysRevLett.38.1440}}.

\bibitem[Wilczek(1978)]{Wilczek:1977pj}
Wilczek, F.
\newblock {Problem of Strong $P$ and $T$ Invariance in the Presence of
  Instantons}.
\newblock {\em Phys. Rev. Lett.} {\bf 1978}, {\em 40},~279--282.
\newblock {\url{https://doi.org/10.1103/PhysRevLett.40.279}}.

\bibitem[Weinberg(1978)]{Weinberg:1977ma}
Weinberg, S.
\newblock {A New Light Boson?}
\newblock {\em Phys. Rev. Lett.} {\bf 1978}, {\em 40},~223--226.
\newblock {\url{https://doi.org/10.1103/PhysRevLett.40.223}}.

\bibitem[Adams et~al.(2022)]{Adams:2022pbo}
Adams, C.B.;  et~al.
\newblock {Axion Dark Matter}.
\newblock In Proceedings of the {Snowmass 2021},  3 2022,
  \href{http://arxiv.org/abs/2203.14923}{{\normalfont
  [arXiv:hep-ex/2203.14923]}}.

\bibitem[Alemany~Fern\'andez et~al.(2025)]{PBC:2025sny}
Alemany~Fern\'andez, R.;  et~al.
\newblock {Summary Report of the Physics Beyond Colliders Study at CERN} {\bf
  2025}.
\newblock  \href{http://arxiv.org/abs/2505.00947}{{\normalfont
  [arXiv:hep-ex/2505.00947]}}.

\bibitem[Banerjee et~al.(2021)Banerjee, Bernhard, Brugger, Charitonidis, Doble,
  Gatignon, and Gerbershagen]{Banerjee:2774716}
Banerjee, D.; Bernhard, J.; Brugger, M.; Charitonidis, N.; Doble, N.; Gatignon,
  L.; Gerbershagen, A.
\newblock {The North Experimental Area at the Cern Super Proton Synchrotron}
  {\bf 2021}.
\newblock Dedicated to Giorgio Brianti on the 50th anniversary of his founding
  the SPS Experimental Areas Group of CERN-Lab II and hence initiating the
  present Enterprise., {\url{https://doi.org/10.17181/CERN.GP3K.0S1Y}}.

\bibitem[Crivelli(2020)]{cernnews}
Crivelli, P.
\newblock Closing the gap between beam dump and colliders: ALP searches with
  NA64.
\newblock
  \url{https://ep-news.web.cern.ch/content/closing-gap-between-beam-dump-and-colliders-alps-searches-na64},
   2020.

\bibitem[Andreev et~al.(2025)]{NA64:2025ddk}
Andreev, Y.M.;  et~al.
\newblock {Searching for Light Dark Matter and Dark Sectors with the NA64
  experiment at the CERN SPS} {\bf 2025}.
\newblock  \href{http://arxiv.org/abs/2505.14291}{{\normalfont
  [arXiv:hep-ex/2505.14291]}}.

\bibitem[Bjorken et~al.(1988)Bjorken, Ecklund, Nelson, Abashian, Church, Lu,
  Mo, Nunamaker, and Rassmann]{Bjorken:1988as}
Bjorken, J.D.; Ecklund, S.; Nelson, W.R.; Abashian, A.; Church, C.; Lu, B.; Mo,
  L.W.; Nunamaker, T.A.; Rassmann, P.
\newblock {Search for Neutral Metastable Penetrating Particles Produced in the
  SLAC Beam Dump}.
\newblock {\em Phys. Rev. D} {\bf 1988}, {\em 38},~3375.
\newblock {\url{https://doi.org/10.1103/PhysRevD.38.3375}}.

\bibitem[Barbiellini(1982)]{barbiellini1982sps}
Barbiellini, G.
\newblock Search for Massive Neutrinos, Axions and SUSY Particles at CERN SPS
  Energies.
\newblock In Proceedings of the Workshop on SPS Fixed Target Physics for the
  Years 1984--1989,  1982.
\newblock \url{https://cds.cern.ch/record/147478}.

\bibitem[Bergsma et~al.(1985)]{CHARM:1985anb}
Bergsma, F.;  et~al.
\newblock {Search for Axion Like Particle Production in 400-{GeV} Proton -
  Copper Interactions}.
\newblock {\em Phys. Lett. B} {\bf 1985}, {\em 157},~458--462.
\newblock {\url{https://doi.org/10.1016/0370-2693(85)90400-9}}.

\bibitem[Donnelly et~al.(1978)Donnelly, Freedman, Lytel, Peccei, and
  Schwartz]{Donnelly:1978ty}
Donnelly, T.W.; Freedman, S.J.; Lytel, R.S.; Peccei, R.D.; Schwartz, M.
\newblock {Do Axions Exist?}
\newblock {\em Phys. Rev. D} {\bf 1978}, {\em 18},~1607.
\newblock {\url{https://doi.org/10.1103/PhysRevD.18.1607}}.

\bibitem[D\"obrich et~al.(2019)D\"obrich, Jaeckel, and
  Spadaro]{Dobrich:2019dxc}
D\"obrich, B.; Jaeckel, J.; Spadaro, T.
\newblock {Light in the beam dump - ALP production from decay photons in proton
  beam-dumps}.
\newblock {\em JHEP} {\bf 2019}, {\em 05},~213,
  \href{http://arxiv.org/abs/1904.02091}{{\normalfont
  [arXiv:hep-ph/1904.02091]}}.
\newblock [Erratum: JHEP 10, 046 (2020)],
  {\url{https://doi.org/10.1007/JHEP05(2019)213}}.

\bibitem[Riordan et~al.(1987)]{Riordan:1987aw}
Riordan, E.M.;  et~al.
\newblock {A Search for Short Lived Axions in an Electron Beam Dump
  Experiment}.
\newblock {\em Phys. Rev. Lett.} {\bf 1987}, {\em 59},~755.
\newblock {\url{https://doi.org/10.1103/PhysRevLett.59.755}}.

\bibitem[Bross et~al.(1991)Bross, Crisler, Pordes, Volk, Errede, and
  Wrbanek]{Bross:1989mp}
Bross, A.; Crisler, M.; Pordes, S.H.; Volk, J.; Errede, S.; Wrbanek, J.
\newblock {A Search for Shortlived Particles Produced in an Electron Beam
  Dump}.
\newblock {\em Phys. Rev. Lett.} {\bf 1991}, {\em 67},~2942--2945.
\newblock {\url{https://doi.org/10.1103/PhysRevLett.67.2942}}.

\bibitem[Badier et~al.(1986)]{NA3:1986ahv}
Badier, J.;  et~al.
\newblock {Mass and Lifetime Limits on New Longlived Particles in 300-{GeV}/$c
  \pi^-$ Interactions}.
\newblock {\em Z. Phys. C} {\bf 1986}, {\em 31},~21.
\newblock {\url{https://doi.org/10.1007/BF01559588}}.

\bibitem[Blumlein et~al.(1991)]{Blumlein:1990ay}
Blumlein, J.;  et~al.
\newblock {Limits on neutral light scalar and pseudoscalar particles in a
  proton beam dump experiment}.
\newblock {\em Z. Phys. C} {\bf 1991}, {\em 51},~341--350.
\newblock {\url{https://doi.org/10.1007/BF01548556}}.

\bibitem[Gninenko(2000)]{Gninenko:2000ds}
Gninenko, S.N.
\newblock {Search for eV (pseudo)scalar penetrating particles in the SPS
  neutrino beam}.
\newblock {\em Nucl. Phys. B Proc. Suppl.} {\bf 2000}, {\em 87},~105--107.
\newblock {\url{https://doi.org/10.1016/S0920-5632(00)00646-0}}.

\bibitem[Gninenko(2014)]{Gninenko:2013rka}
Gninenko, S.N.
\newblock {Search for MeV dark photons in a light-shining-through-walls
  experiment at CERN}.
\newblock {\em Phys. Rev. D} {\bf 2014}, {\em 89},~075008,
  \href{http://arxiv.org/abs/1308.6521}{{\normalfont
  [arXiv:hep-ph/1308.6521]}}.
\newblock {\url{https://doi.org/10.1103/PhysRevD.89.075008}}.

\bibitem[Gninenko et~al.(2016)Gninenko, Krasnikov, Kirsanov, and
  Kirpichnikov]{Gninenko:2016kpg}
Gninenko, S.N.; Krasnikov, N.V.; Kirsanov, M.M.; Kirpichnikov, D.V.
\newblock {Missing energy signature from invisible decays of dark photons at
  the CERN SPS}.
\newblock {\em Phys. Rev. D} {\bf 2016}, {\em 94},~095025,
  \href{http://arxiv.org/abs/1604.08432}{{\normalfont
  [arXiv:hep-ph/1604.08432]}}.
\newblock {\url{https://doi.org/10.1103/PhysRevD.94.095025}}.

\bibitem[Andreev et~al.(2023)]{NA64:2023wbi}
Andreev, Y.M.;  et~al.
\newblock {Search for Light Dark Matter with NA64 at CERN}.
\newblock {\em Phys. Rev. Lett.} {\bf 2023}, {\em 131},~161801,
  \href{http://arxiv.org/abs/2307.02404}{{\normalfont
  [arXiv:hep-ex/2307.02404]}}.
\newblock {\url{https://doi.org/10.1103/PhysRevLett.131.161801}}.

\bibitem[Andreev et~al.(2024)]{NA64:2023ehh}
Andreev, Y.M.;  et~al.
\newblock {Probing light dark matter with positron beams at NA64}.
\newblock {\em Phys. Rev. D} {\bf 2024}, {\em 109},~L031103,
  \href{http://arxiv.org/abs/2308.15612}{{\normalfont
  [arXiv:hep-ex/2308.15612]}}.
\newblock {\url{https://doi.org/10.1103/PhysRevD.109.L031103}}.

\bibitem[Andreev et~al.(2025)]{NA64:2025rib}
Andreev, Y.M.;  et~al.
\newblock {Proof of principle for a light dark matter search with low-energy
  positron beams at NA64} {\bf 2025}.
\newblock  \href{http://arxiv.org/abs/2502.04053}{{\normalfont
  [arXiv:hep-ex/2502.04053]}}.

\bibitem[Andreev et~al.(2024{\natexlab{a}})]{NA64:2024mah}
Andreev, Y.M.;  et~al.
\newblock {Dark-Sector Search via Pion-Produced \ensuremath{\eta} and
  \ensuremath{\eta}' Mesons Decaying Invisibly in the NA64h Detector}.
\newblock {\em Phys. Rev. Lett.} {\bf 2024}, {\em 133},~121803,
  \href{http://arxiv.org/abs/2406.01990}{{\normalfont
  [arXiv:hep-ex/2406.01990]}}.
\newblock {\url{https://doi.org/10.1103/PhysRevLett.133.121803}}.

\bibitem[Andreev et~al.(2024{\natexlab{b}})]{NA64:2024klw}
Andreev, Y.M.;  et~al.
\newblock {First Results in the Search for Dark Sectors at NA64 with the CERN
  SPS High Energy Muon Beam}.
\newblock {\em Phys. Rev. Lett.} {\bf 2024}, {\em 132},~211803,
  \href{http://arxiv.org/abs/2401.01708}{{\normalfont
  [arXiv:hep-ex/2401.01708]}}.
\newblock {\url{https://doi.org/10.1103/PhysRevLett.132.211803}}.

\bibitem[Andreev et~al.(2024{\natexlab{c}})]{NA64:2024nwj}
Andreev, Y.M.;  et~al.
\newblock {Shedding light on dark sectors with high-energy muons at the NA64
  experiment at the CERN SPS}.
\newblock {\em Phys. Rev. D} {\bf 2024}, {\em 110},~112015,
  \href{http://arxiv.org/abs/2409.10128}{{\normalfont
  [arXiv:hep-ex/2409.10128]}}.
\newblock {\url{https://doi.org/10.1103/PhysRevD.110.112015}}.

\bibitem[Dusaev et~al.(2020)Dusaev, Kirpichnikov, and Kirsanov]{Dusaev:2020gxi}
Dusaev, R.R.; Kirpichnikov, D.V.; Kirsanov, M.M.
\newblock {Photoproduction of axionlike particles in the NA64 experiment}.
\newblock {\em Phys. Rev. D} {\bf 2020}, {\em 102},~055018,
  \href{http://arxiv.org/abs/2004.04469}{{\normalfont
  [arXiv:hep-ph/2004.04469]}}.
\newblock {\url{https://doi.org/10.1103/PhysRevD.102.055018}}.

\bibitem[Banerjee et~al.(2020)]{NA64:2020qwq}
Banerjee, D.;  et~al.
\newblock {Search for Axionlike and Scalar Particles with the NA64 Experiment}.
\newblock {\em Phys. Rev. Lett.} {\bf 2020}, {\em 125},~081801,
  \href{http://arxiv.org/abs/2005.02710}{{\normalfont
  [arXiv:hep-ex/2005.02710]}}.
\newblock {\url{https://doi.org/10.1103/PhysRevLett.125.081801}}.

\bibitem[Dine et~al.(1981)Dine, Fischler, and Srednicki]{Dine:1981rt}
Dine, M.; Fischler, W.; Srednicki, M.
\newblock {A Simple Solution to the Strong CP Problem with a Harmless Axion}.
\newblock {\em Phys. Lett. B} {\bf 1981}, {\em 104},~199--202.
\newblock {\url{https://doi.org/10.1016/0370-2693(81)90590-6}}.

\bibitem[Kim(1979)]{Kim:1979if}
Kim, J.E.
\newblock {Weak Interaction Singlet and Strong CP Invariance}.
\newblock {\em Phys. Rev. Lett.} {\bf 1979}, {\em 43},~103.
\newblock {\url{https://doi.org/10.1103/PhysRevLett.43.103}}.

\bibitem[Tanabashi et~al.(2018)]{ParticleDataGroup:2018ovx}
Tanabashi, M.;  et~al.
\newblock {Review of Particle Physics}.
\newblock {\em Phys. Rev. D} {\bf 2018}, {\em 98},~030001.
\newblock {\url{https://doi.org/10.1103/PhysRevD.98.030001}}.

\bibitem[Kim(1998)]{Kim:1998va}
Kim, J.E.
\newblock {Constraints on very light axions from cavity experiments}.
\newblock {\em Phys. Rev. D} {\bf 1998}, {\em 58},~055006,
  \href{http://arxiv.org/abs/hep-ph/9802220}{{\normalfont [hep-ph/9802220]}}.
\newblock {\url{https://doi.org/10.1103/PhysRevD.58.055006}}.

\bibitem[Anelli et~al.(2005)]{Anelli:2005ju}
Anelli, G.;  et~al.
\newblock {Proposal to measure the rare decay $K^+ \rightarrow \pi^+ \nu
  \overline{\nu}$ at the CERN SPS} {\bf 2005}.

\bibitem[Cortina~Gil et~al.(2017)]{NA62:2017rwk}
Cortina~Gil, E.;  et~al.
\newblock {The Beam and detector of the NA62 experiment at CERN}.
\newblock {\em JINST} {\bf 2017}, {\em 12},~P05025,
  \href{http://arxiv.org/abs/1703.08501}{{\normalfont
  [arXiv:physics.ins-det/1703.08501]}}.
\newblock {\url{https://doi.org/10.1088/1748-0221/12/05/P05025}}.

\bibitem[Beacham et~al.(2020)]{Beacham:2019nyx}
Beacham, J.;  et~al.
\newblock {Physics Beyond Colliders at CERN: Beyond the Standard Model Working
  Group Report}.
\newblock {\em J. Phys. G} {\bf 2020}, {\em 47},~010501,
  \href{http://arxiv.org/abs/1901.09966}{{\normalfont
  [arXiv:hep-ex/1901.09966]}}.
\newblock {\url{https://doi.org/10.1088/1361-6471/ab4cd2}}.

\bibitem[Cortina~Gil et~al.(2025)]{NA62:2025yzs}
Cortina~Gil, E.;  et~al.
\newblock {Search for hadronic decays of feebly-interacting particles at NA62}.
\newblock {\em Eur. Phys. J. C} {\bf 2025}, {\em 85},~571,
  \href{http://arxiv.org/abs/2502.04241}{{\normalfont
  [arXiv:hep-ex/2502.04241]}}.
\newblock {\url{https://doi.org/10.1140/epjc/s10052-025-14133-w}}.

\bibitem[Cortina~Gil et~al.(2021{\natexlab{a}})]{NA62:2020pwi}
Cortina~Gil, E.;  et~al.
\newblock {Search for $\pi^0$ decays to invisible particles}.
\newblock {\em JHEP} {\bf 2021}, {\em 02},~201,
  \href{http://arxiv.org/abs/2010.07644}{{\normalfont
  [arXiv:hep-ex/2010.07644]}}.
\newblock {\url{https://doi.org/10.1007/JHEP02(2021)201}}.

\bibitem[Cortina~Gil et~al.(2021{\natexlab{b}})]{NA62:2020xlg}
Cortina~Gil, E.;  et~al.
\newblock {Search for a feebly interacting particle $X$ in the decay
  $K^{+}\rightarrow\pi^{+}X$}.
\newblock {\em JHEP} {\bf 2021}, {\em 03},~058,
  \href{http://arxiv.org/abs/2011.11329}{{\normalfont
  [arXiv:hep-ex/2011.11329]}}.
\newblock {\url{https://doi.org/10.1007/JHEP03(2021)058}}.

\bibitem[Cortina~Gil et~al.(2024)]{NA62:2023olg}
Cortina~Gil, E.;  et~al.
\newblock {Measurement of the $K^+\rightarrow{}\pi^+\gamma\gamma$ decay}.
\newblock {\em Phys. Lett. B} {\bf 2024}, {\em 850},~138513,
  \href{http://arxiv.org/abs/2311.01837}{{\normalfont
  [arXiv:hep-ex/2311.01837]}}.
\newblock {\url{https://doi.org/10.1016/j.physletb.2024.138513}}.

\bibitem[Banerjee et~al.(2020)]{NA64:2019auh}
Banerjee, D.;  et~al.
\newblock {Improved limits on a hypothetical X(16.7) boson and a dark photon
  decaying into $e^+e^-$ pairs}.
\newblock {\em Phys. Rev. D} {\bf 2020}, {\em 101},~071101,
  \href{http://arxiv.org/abs/1912.11389}{{\normalfont
  [arXiv:hep-ex/1912.11389]}}.
\newblock {\url{https://doi.org/10.1103/PhysRevD.101.071101}}.

\bibitem[Depero et~al.(2020)]{NA64:2020xxh}
Depero, E.;  et~al.
\newblock {Hunting down the X17 boson at the CERN SPS}.
\newblock {\em Eur. Phys. J. C} {\bf 2020}, {\em 80},~1159,
  \href{http://arxiv.org/abs/2009.02756}{{\normalfont
  [arXiv:hep-ex/2009.02756]}}.
\newblock {\url{https://doi.org/10.1140/epjc/s10052-020-08725-x}}.

\bibitem[Li et~al.(2025)Li, Liu, and Song]{Li:2025yzb}
Li, H.; Liu, Z.; Song, N.
\newblock {Probing axion and muon-philic new physics with muon beam dump} {\bf
  2025}.
\newblock  \href{http://arxiv.org/abs/2501.06294}{{\normalfont
  [arXiv:hep-ph/2501.06294]}}.

\bibitem[Eberhart et~al.(2025)Eberhart, Fedele, Kahlhoefer, Ravensburg, and
  Ziegler]{Eberhart:2025lyu}
Eberhart, A.; Fedele, M.; Kahlhoefer, F.; Ravensburg, E.; Ziegler, R.
\newblock {Leptophilic ALPs in Laboratory Experiments} {\bf 2025}.
\newblock  \href{http://arxiv.org/abs/2504.05873}{{\normalfont
  [arXiv:hep-ph/2504.05873]}}.

\bibitem[Jerhot(2025)]{Jerhot:2936260}
Jerhot, J.
\newblock {Projection of finalized NA62 beam-dump analyses to the full
  2021-2026 statistics} {\bf 2025}.

\bibitem[Mammen~Abraham et~al.(2025)]{FASER:2024bbl}
Mammen~Abraham, R.;  et~al.
\newblock {Shining light on the dark sector: search for axion-like particles
  and other new physics in photonic final states with FASER}.
\newblock {\em JHEP} {\bf 2025}, {\em 01},~199,
  \href{http://arxiv.org/abs/2410.10363}{{\normalfont
  [arXiv:hep-ex/2410.10363]}}.
\newblock {\url{https://doi.org/10.1007/JHEP01(2025)199}}.

\bibitem[Feng et~al.(2018)Feng, Galon, Kling, and Trojanowski]{Feng:2017uoz}
Feng, J.L.; Galon, I.; Kling, F.; Trojanowski, S.
\newblock {ForwArd Search ExpeRiment at the LHC}.
\newblock {\em Phys. Rev. D} {\bf 2018}, {\em 97},~035001,
  \href{http://arxiv.org/abs/1708.09389}{{\normalfont
  [arXiv:hep-ph/1708.09389]}}.
\newblock {\url{https://doi.org/10.1103/PhysRevD.97.035001}}.

\bibitem[Ahdida et~al.(2022)]{SHiP:2021nfo}
Ahdida, C.;  et~al.
\newblock {The SHiP experiment at the proposed CERN SPS Beam Dump Facility}.
\newblock {\em Eur. Phys. J. C} {\bf 2022}, {\em 82},~486,
  \href{http://arxiv.org/abs/2112.01487}{{\normalfont
  [arXiv:physics.ins-det/2112.01487]}}.
\newblock {\url{https://doi.org/10.1140/epjc/s10052-022-10346-5}}.

\end{thebibliography}

\end{document}